\documentclass[aps, prd, twocolumn, amsmath, floats,floatfix, superscriptaddress, nofootinbib, showkeys]{revtex4}
\usepackage{amsmath} 
\usepackage{amsfonts,hyperref}
\usepackage{slashed}
\usepackage[dvips]{graphicx}
\usepackage{color}
\usepackage{units}
\usepackage{bbm}
\usepackage{soul}
\usepackage{mathrsfs}
\allowdisplaybreaks 

\def\be{\begin{equation}}
\def\ee{\end{equation}}
\def\bea{\begin{eqnarray}}
\def\eea{\end{eqnarray}}

\begin{document}

\title{Energy condition and cosmic censorship conjecture in the perfect fluid collapse}

\author{Javad T. Firouzjaee}

\affiliation{Department of Physics, K. N. Toosi University of Technology, \\
	P. O. Box 15875-4416, Tehran, Iran}
\affiliation{ School of Physics, Institute for Research in Fundamental Sciences (IPM), \\
	P. O. Box 19395-5531, Tehran, Iran } 
\email{firouzjaee@kntu.ac.ir}

\begin{abstract}
	The hypothesis of cosmic censorship plays a crucial role in classical general relativity, namely, to ensure that naked singularities would never occur from the black hole singularity. In this paper, we will present how energy conditions prohibit forming the naked singularity in the spherical perfect fluid collapse, and thus the strong cosmic censorship conjecture hold in this model. We also show that this result can be extended to the cosmological constant background.
\end{abstract}

\date{\today}

\maketitle

\section{Introduction}

Hawking and Penrose have shown spacetime singularities are unavoidable when the trapped region form in general relativity and suitable energy conditions are satisfied \cite{HawPen}. The key issue is whether these singularities are covered by the event horizon as in the Schwarzschild case or not, that is whether they are black holes \textit{or naked singularities}. Although Penrose’s cosmic censorship conjecture (CCC) \cite{Penrose} states that physically reasonable initial data cannot end in naked singularities visible to distant observers, various counterexamples have been proposed. The issue is whether these are astrophysically plausible counterexamples to the CCC.  These counterexamples usually challenged the strong cosmic censorship conjecture that, asserts that, generically, timelike singularities never occur, so that even an observer who falls into a black hole will never “see” the singularity. In counterpoint, when singularities (or locally naked singularity) of gravitational collapse are contained in black holes is known as the weak cosmic censorship conjecture \cite{Wald:1997wa}. 

Indeed CCC counterexamples have been proposed in the case of pressure-free spherically symmetric (LTB) collapse, which have  attracted major interest \cite{LTB-naked,LTB-naked2,LTB-naked3} (see \cite{Joshi-book-review} for review).  It seems that naked singularity could occur in cases where there is no pressure, and although pressure will occur and could alter the situation in more realistic cases, these examples provide useful toy models as a basis for further investigation.  
Most of the attempts are focused on finding timelike singularities from which the null geodesics emerge (at least locally) \cite{other-ref}. In this letter, we will present how energy conditions prohibit forming the naked singularity.

Our approach to show the CCC in the perfect fluid collapse is based on the fact that the apparent horizon which is coincident with the singularity at the center cannot be future timelike.  We assume that the weak energy condition, which states that the energy density is not negative, is satisfied and will see how this is playing an important role in this context. In the following, we use units in
which $ G = c = 1 $ and our reasoning is based on the radial null geodesics.\\

\section{Perfect fluid spherical collapse}\label{sec:fluid}

A collapsing perfect fluid within a compact spherically symmetric
spacetime region can be described  \cite{Misner:1964je} by the following metric in the comoving
coordinates $(t,r,\theta,\varphi)$:
\bea
ds^{2}&=&-e^{2\nu(t,r)}dt^{2}+e^{2\psi(t,r)}dr^{2}+R(t,r)^{2}d\Omega^{2}, \, \nonumber \\ \,u^\mu &=& e^{-\nu(t,r)}\delta^\mu_0.
\eea
Assume the energy momentum tensor has the  perfect fluid form
\begin{eqnarray}
T^{t}_{t}=-\rho(t,r), 
\,\, 
T^{r}_{r} = T^{\theta}_{\theta}=T^{\varphi}_{\varphi}= 
p(t,r). 
\end{eqnarray}
The Einstein equations for this energy momentum tensor give,
\bea \label{gltbe22}
\rho(t,r)=\frac{2M'(t,r)}{R^{2}(t,r)R'(t,r)},\nonumber 
\\ ~~p(t,r)=-\frac{2\dot{M(t,r)}}{R^{2}(t,r)\dot{R}(t,r)},
\eea
\begin{equation}
\nu'(t,r)= 
-\frac{p'(t,r) 
}{\rho(t,r)+p(t,r) 
},
\end{equation}
where $\dot{}$ and $'$ mean the partial derivative relative to $t$ and $r$ respectively. Four velocity of this comoving fluid is $u^\mu=(e^{-\nu},0,0,0)$ and after contraction this vector with the energy momentum tensor, $\nabla_\nu T^{\mu \nu}=0$, we get the energy-momentum conservation in the form
\be
\dfrac{d\rho}{d\tau}+ \theta (\rho+p)= 0,
\ee
where $\theta = \nabla_\mu u^\mu$ and $\dfrac{d}{d\tau}=e^{-\nu} \dfrac{d}{dt}$. As a result, one gets
\bea
e^{-\nu}\dfrac{d\rho}{dt}+ (\dot{\nu}+\dot{\psi}+2\dfrac{\dot{R}}{R}) (\rho+p) =0 \nonumber \\
-2\dot{R}'+R'\frac{\dot{G}}{G}+\dot{R}\frac{H'}{H}=0,
\eea
where
\begin{equation}
\label{G H}
G=e^{-2\psi}(R')^{2}~~,~~H=e^{-2\nu}(\dot{R})^{2},
\end{equation}
and $M(t,r)$ quantity is defined as
\bea 
\label{gltbe3} G(t,r)-H(t,r)=1-\frac{2M(t,r)}{R(t,r)}.
\eea
It is obvious from the equation \eqref{G H} that  $G, H > 0$ are non-negative quantities. It is known that the apparent horizon  (that is
the boundary of the trapped region) is located at $R=2M$ \cite{Hayward:1994bu,Firouzjaee-penn}. The $ M $ quantity is the Misner-Sharp Energy 
\begin{eqnarray} \label{ms-def}
E_{MS}=M(R)=\frac{1}{2}\int_{0}^{R}\rho R^{2}dR,
\end{eqnarray}
or
\begin{equation}
E_{MS}=\frac{1}{8\pi}\int_{0}^{r}\rho
\sqrt{(1+(\frac{dR}{d\tau})^{2}-\frac{2M}{R})}d^{3}V,
\end{equation}
where
$ d^{3}V=4\pi e^{\psi}R'dr,$ and $\frac{d}{d\tau}=e^{-\nu}\frac{d}{dt}.$
The last form of the function $M$ indicates that when considered as energy, it includes contribution terms the kinetic energy and the gravitational potential energy.  \\

Equation \eqref{ms-def} shows that non-singular density gives $M(R\rightarrow0)=0$. This equation can be held just before the singularity formation (onset of singularity formation) when $\rho/\rho_0 \gg 1 $ where $\rho_0 $ can be the average density of collapsing object.

During the gravitational collapse of a cloud of matter, a singularity will form when physical quantities like the energy density $\rho$ and the Ricci scalar $R$ will diverge.
There can be two kinds of singularities in this metric: a big bang (big crunch) singularity (shell-focusing singularity) 
at $R(t,r)=0$, and a shell-crossing one at $R'(t,r)=0$. 
\textit{Shell-crossing} singularities are gravitationally weak which there are proposals for extending the
spacetime through such singularities and show a breakdown of the coordinate system being used, rather than a genuine physical singularity. These singularities are generally not considered as being serious counterexamples to the cosmic censorship conjecture \cite{Shell-crossing-coordinate}. Indeed it has been suggested one need not consider 
the shell-crossing singularity because even if they occur in the dust case, they are likely to not occur in the fluid case 
because of pressure effects. To show that the shell-crossing singularity cannot be naked, suppose a shell-crossing singularity first forms at $R_0 (t_{sc}) \neq 0$. To examine whether a point is in the trapped surface or not, first, we must calculate the ratio $\frac{R}{2M}$ to see that it is greater than one or not. If $\frac{R}{2M} <1$ this point will be trapped and no null or timelike geodesic can escape from it to infinity. Consider a sphere with radius $R_0+\epsilon$ which the $\epsilon$ can has any small positive value before the  shell-crossing singularity form $t<t_{sc}$. The ratio of $\frac{R}{2M}$ for this point gets
\be
\frac{R}{2M}=\lim_{R \rightarrow R_0} \frac{R}{8 \pi \int_0^{R+\epsilon} \rho R^2 dR}.
\ee
Here, we wrote the Misner-Sharp mass in the $(t, R)$ coordinate. After Taylor expansion of the above ratio around the $R_0$ you'll get, $\frac{R}{2M} \sim \frac{1}{8 \pi \rho R_0^2 } <1$, which is clearly less than $1$ as density goes to infinity at $R=R_0$. As a result, the trapped surface will form just before the shell-crossing singularity formation and outside it. Consequently, no naked singularity forms from these shell-crossing singularities.\\

Note that in contrast to the dust collapse which always ends in a singularity at the center \cite{Firouzjaee-2010}, the perfect fluid collapse can be stopped by pressure at the center and the collapse ends to a compact star, not a black hole \cite{Moradi:2015caa}. On the apparent horizon, equation (\ref{gltbe3}) gives $G=H$.
The apparent horizon slope in the $(t,r)$ plane is
\begin{equation}
\label{AHdtdr}
\frac{dt}{dr}|_{AH}=\frac{R' -2 M'}{2\dot{M}-\dot{R}},
\end{equation}

The tangent vector on the apparent horizon can be presented as
\begin{equation} 
k^\mu=c(1,\frac{dr}{dt},0,0),
\end{equation}
where $c$ is a constant.
To see whether this tangent vector is timelike or spacelike, we have to calculate:
\begin{equation} 
k_\mu k^\mu= c^2 (-e^{2\nu}+e^{2\psi} (\frac{dr}{dt})^2). 
\end{equation}

Using the equation \eqref{AHdtdr}, we have
\begin{equation} 
k_\mu k^\mu= c^2 e^{2\nu}(-1+\frac{e^{2\psi}}{e^{2\nu}} (\frac{2\dot{M}-\dot{R}}{R'-2M'})^2)	.	
\end{equation}
Since on the apparent horizon, we have $G=H \rightarrow \frac{e^{2\psi}}{e^{2\nu}}=\frac{R'^2}{\dot{R}^2}$, and applying the Einstein equation (\ref{gltbe22}) we get $ 2 \frac{M'}{R'} = \rho R^2 >0 $ and $ 2 \frac{\dot{M}}{\dot{R}}=- p R^2 >0 $. Using these equations, the above equation will be simplified as
\begin{equation} 
k_\mu k^\mu= c^2 e^{2\nu} (\frac{(1+pR^2)^2}{(1-\rho R^2)^2}-1).
\end{equation}

If the weak energy conditions, $ \rho \ge 0 $ and $ \rho+ p > 0 $, are held in he model, then we get that $k_\mu k^\mu >1$. Thus this condition does not allow the apparent horizon to become future timelike. This argument does not alter with changing the coordinate because the causal nature of spacetime does not change by using another coordinate.\\

To have a more relativistic model, consider a black hole which is formed in dark energy background with the cosmological constant. If we solve the Einstein equation with the spherical symmetric metric and the perfect fluid energy momentum tensor, we get 
\bea \label{gltbe22-L}
\rho(t,r)=\frac{2M'(t,r)}{R^{2}(t,r)R'(t,r)}-\Lambda,\nonumber 
\\ ~~p(t,r)=-\frac{2\dot{M(t,r)}}{R^{2}(t,r)\dot{R}(t,r)}+\Lambda,
\eea

where $M(t,r)$ quantity is defined  in the following equation
\bea 
\label{gltbe3-L} 
G(t,r)-H(t,r)=1-\frac{2M(t,r)}{R(t,r)}+\frac{\Lambda R(t,r)^2}{3}.
\eea
The  $ M(R) $ quantity is the Misner-Sharp Energy 
\begin{eqnarray} \label{ms-def-L}
E_{MS}=\frac{1}{2}\int_{0}^{R}\rho R^{2}dR,
\end{eqnarray}
or
\begin{equation}
E_{MS}=\frac{1}{8\pi}\int_{0}^{r}\rho
\sqrt{(1+(\frac{dR}{d\tau})^{2}-\frac{2M}{R}+\frac{\Lambda R^2}{3})}d^{3}V.
\end{equation}
As discussed in \cite{Firouzjaee-penn}, the cosmological constant contributes to the Misner-Sharp Energy, but we can discriminate the matter-energy in a comoving sphere with radius $r$ and time $t$ with $M(t,r)$. The apparent horizon location will be changed by cosmological constant as $R=2M-\frac{\Lambda R^3}{3}$. This equation shows that the positive cosmological constant will reduce the horizon radius for a given matter mass, $M$. Nevertheless, using the present cosmological constant value for the astrophysical black hole in this equation leads to negligible correction for the astrophysical apparent horizon radius. \\
In the presence of the cosmological constant, the apparent horizon slope in the $(t,r)$ plane is
\begin{equation}
\label{AHdtd-L}
\frac{dt}{dr}|_{AH}=\frac{R' -2 M'+\Lambda R^2 R'}{2\dot{M}-\dot{R}- \Lambda R^2 \dot{R}},
\end{equation}

To see whether this tangent vector is timelike or spacelike in presence of cosmological constant, we have to calculate the magnitude of the apparent horizon tangent vector:
\begin{equation} 
k_\mu k^\mu= c^2 (-e^{2\nu}+e^{2\psi} (\frac{dr}{dt})^2). 
\end{equation}

Now, using the equation \eqref{AHdtd-L}
\begin{equation} 
k_\mu k^\mu= c^2 e^{2\nu}(-1+\frac{e^{2\psi}}{e^{2\nu}} (\frac{2\dot{M}-\dot{R}-\dot{R}- \Lambda R^2 \dot{R}}{R'-2M'-\dot{R}+ \Lambda R^2 R'})^2)	.	
\end{equation}
On the apparent horizon, we have $G=H \rightarrow \frac{e^{2\psi}}{e^{2\nu}}=\frac{R'^2}{\dot{R}^2}$, and using the Einstein equation (\ref{gltbe22-L}) we get $ 2 \frac{M'}{R'} = \rho R^2+\Lambda R^2 >0 $ and $ 2 \frac{\dot{M}}{\dot{R}}=- p R^2 +\Lambda R^2>0 $. Applying these equations to the magnitude of the apparent horizon tangent vector
\begin{equation} 
k_\mu k^\mu= c^2 e^{2\nu} (\frac{(1+pR^2)^2}{(1-\rho R^2)^2}-1).
\end{equation}

It is clear from the weak energy condition, $ \rho \ge 0 $ and $ \rho+ p > 0 $, that $k_\mu k^\mu >1$, thus this condition does not allow the apparent horizon to become future timelike.

\begin{figure}[]
	\begin{center}
		\includegraphics[width=3.3in]{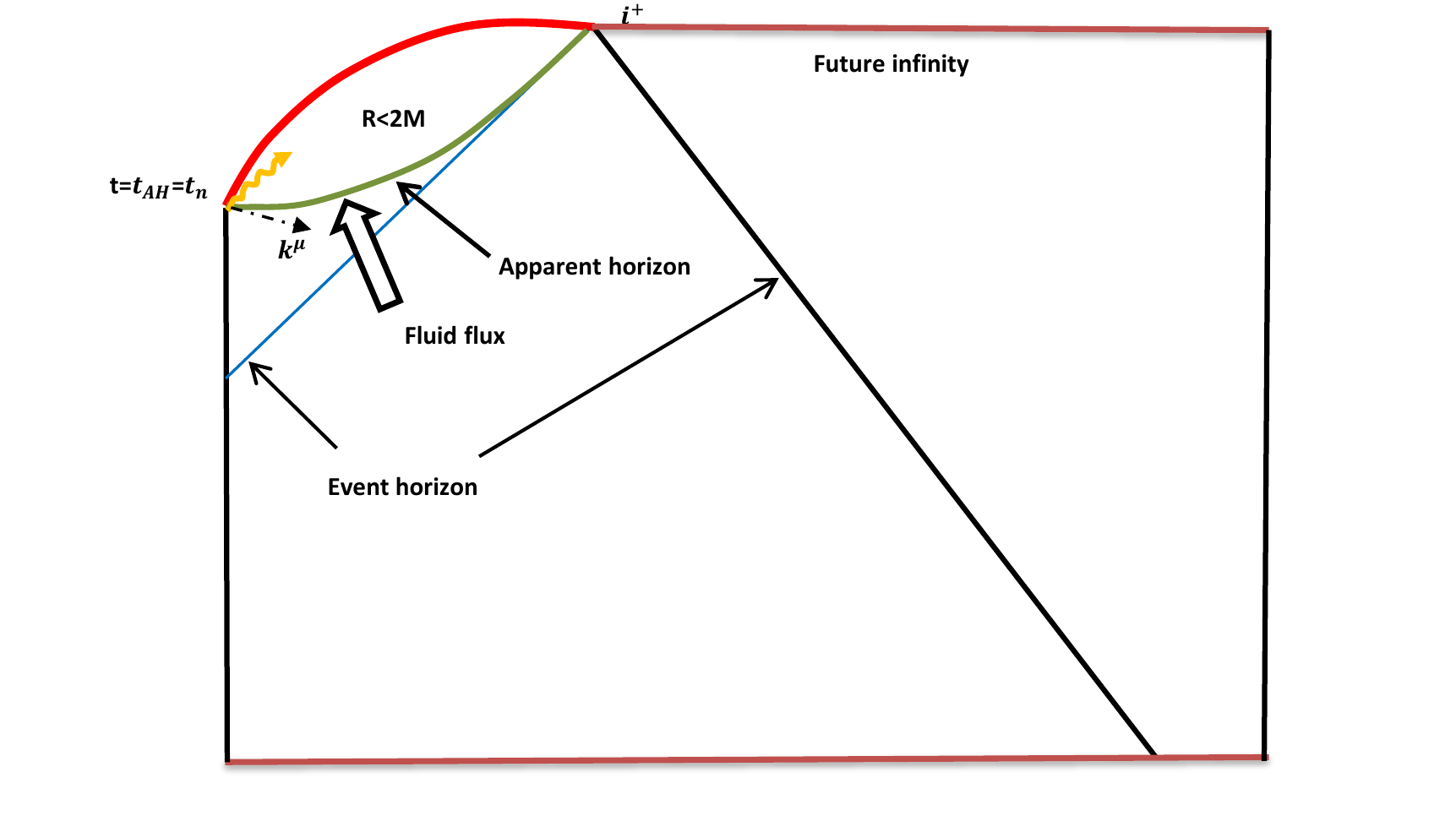}
		\caption{  Penrose diagram of the fluid collapse in the cosmological de Sitter background.}
		\label{Penrose}
	\end{center}
\end{figure}

As shown in   Figure (\ref{Penrose}), the only point which has a chance to be a naked singularity is at the center of the black hole.  Since the apparent horizon cannot be future timelike, then no light can escape from this point of the singularity, and therefore no naked singularity happens in the perfect fluid collapse which has the central singularity $R=0$. 

\begin{figure}[]
	\begin{center}
		\includegraphics[width=3in]{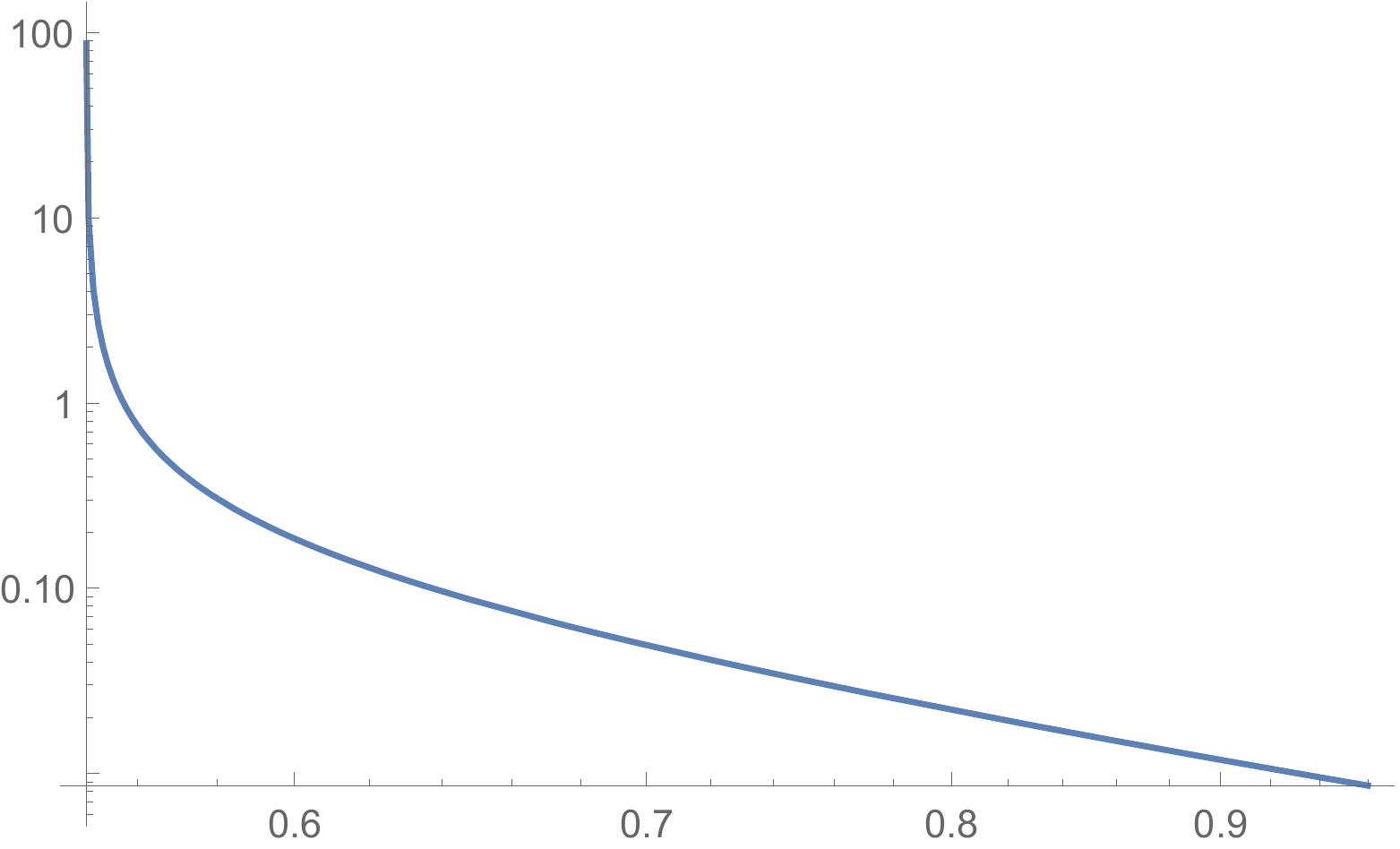}
		\caption{ Density profile of collapse at onset of singularity formation.}
		\label{denfig}
	\end{center}
\end{figure}

\begin{figure}[]
	\begin{center}
		\includegraphics[width=3in]{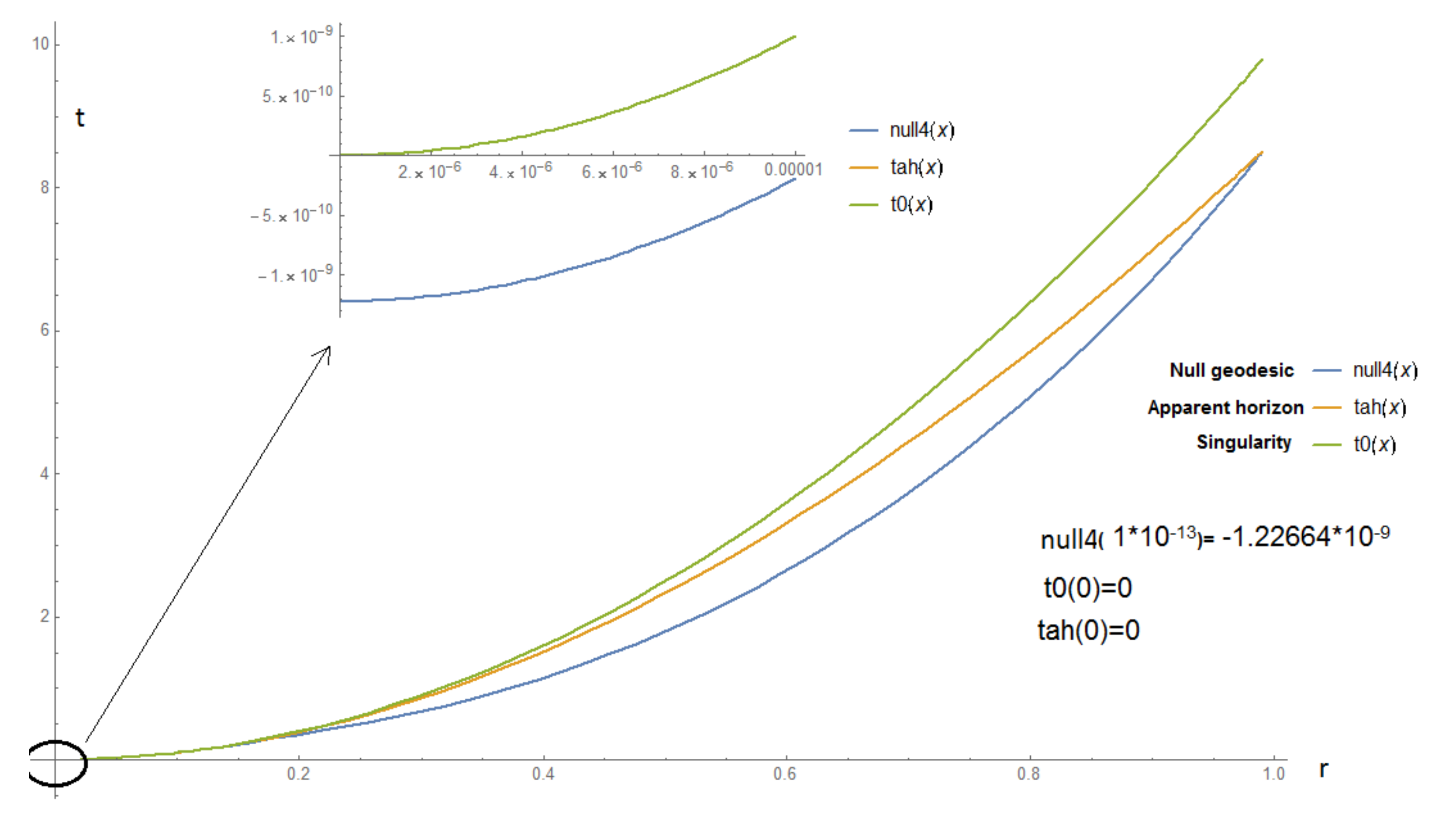}
		\caption{Null geodesic behaviour near singularity formation.}
		\label{null.sin.fig}
	\end{center}
\end{figure}

In this part, we consider the Eardely and Smarr model \cite{LTB-naked} which is claimed that contains the naked singularity. They consider a flat LTB model whose mass function is in the form $M(r)=r^3$ and the singularity time for each layer is $t_0(r)=\zeta r^p$.
In Figure (\ref{denfig}), the density profile of collapse at the onset of singularity formation for the parameters, $p=2$ and
$\zeta=10$ is depicted. In Figure (\ref{null.sin.fig}), a typical null geodesics after $t>0$ (before singularity formation) near the singularity are shown. These three geodesics cross the apparent horizon and end to the singularity.
This diagram verifies some points:
\begin{itemize}
	\item The apparent horizon is spacelike.
	\item Singularity is spacelike.
	\item This behavior will be held even in the case that the null geodesic is close to the crossing point i.e $R=M=0$.
	\item Another reason for holding the CCC is that we see that a typical null geodesic near the singularity point will fall into the trapped region (and then into a singularity). Since null geodesic congruences do not cross each other then all null geodesics originated from the center and after this null geodesics will also fall into the trapped region and then it holds CCC.
\end{itemize}
These confirm that no naked singularity occurs in this model.

\emph{\textbf{Conclusion:}} 

The main issue of the present investigation was to examine whether the end state of the collapse is a naked singularity or a black hole in the perfect fluid collapse. Our results show that having the weak energy condition causes the slope of the apparent horizon can not be future timelike.  As a result, no local naked singularity can occur in the spherical perfect fluid collapse. This feature will not change if the black holes are located in the cosmological constant background. This result will be held by dust (LTB model) collapse which is a special case of the perfect fluid.
As an illustrative example, we took a known model in dust collapse and numerically depict its horizon and singularity curve, and it was shown how a typical null geodesic from the central singularity is trapped.
The above result provides an intriguing perspective on black hole physics which tells that not only do energy conditions play an important role in forming the singularity but also cover it by forming the apparent horizon. All these results lead to this point that the strong cosmic censorship conjecture holds in the perfect fluid spherical collapse.\\


\end{document}